\documentclass[aps,pra,twocolumn,reprint,superscriptaddress]{revtex4-2}
\usepackage{graphicx}
\usepackage{amsmath, amssymb, amsfonts}
\usepackage{xcolor}
\usepackage[colorlinks=true,linkcolor=blue]{hyperref}
\usepackage[normalem]{ulem}
\usepackage{tikz}

\expandafter\def\expandafter\normalsize\expandafter{%
    \normalsize%
    \setlength\abovedisplayskip{1pt}%
    \setlength\belowdisplayskip{1pt}%
}

\begin{document}
\title{Universality in the microwave shielding of ultracold polar molecules}

\author{Joy Dutta}
\affiliation{Joint Quantum Centre (JQC)
Durham-Newcastle, Department of Chemistry, Durham University, South Road,
Durham, DH1 3LE, United Kingdom.}

\author{Bijit Mukherjee}
\affiliation{Faculty of Physics, University of Warsaw, Pasteura 5, 02-093 Warsaw, Poland}

\author{Jeremy M. Hutson}
{\email{j.m.hutson@durham.ac.uk} \affiliation{Joint Quantum Centre (JQC)
Durham-Newcastle, Department of Chemistry, Durham University, South Road,
Durham, DH1 3LE, United Kingdom.}

\date{\today}

\begin{abstract}
Microwave shielding is an important technique that can suppress the losses that arise from collisions of ultracold polar molecules. It has been instrumental in achieving molecular Bose-Einstein condensation (BEC) for NaCs [Bigagli \emph{et al.}, Nature {\bf 631}, 289 (2024)]. We demonstrate that microwave shielding is universal, in the sense that the 2-body collision properties of different molecules are very similar when expressed in suitable reduced units of length and energy. This applies to rate coefficients for inelastic scattering and loss, to scattering lengths, and to the properties of 2-molecule bound states. We also explore the small deviations from universality that arise at very large Rabi frequencies. In general, the collision properties are near-universal except when the Rabi frequency exceeds a few percent of the molecular rotational constant. The universality extends to elliptically polarized microwaves and to combinations of multiple fields. Our results indicate that the methods that have been used to achieve BEC for NaCs can be transferred directly to most other polar molecules.
\end{abstract}

\maketitle

\section{Introduction}
Ultracold polar molecules provide a versatile platform for exploring quantum science. Their applications include quantum simulation \cite{Cornish:2024}, quantum computation \cite{wang:2022}, quantum magnetism \cite{Wall:QMUM:2015}, many-body physics \cite{Baranov:2012}, quantum metrology \cite{DeMille:2024}, and controlled chemical reactions \cite{Karman:2024}.

Production of stable ultracold gases \cite{Langen:2024} necessitates suppression of collisional losses that take place at short range ($R \lesssim 100$~bohr). These losses are mediated by various phenomena including chemical reactions \cite{Ospelkaus:react:2010, Hu:2019}, two-body inelastic loss, three-body recombination \cite{Mayle:2013}, and photoexcitation of collision complexes \cite{Christianen:laser:2019}. For molecular gases confined to two dimensions, they can be suppressed with a strong electric field perpendicular to the plane, producing long-range repulsion between side-by-side dipoles \cite{deMiranda:2011, Valtolina:KRb2D:2020}. This prevents colliding pairs reaching the short-range region where most losses occur. However, this is not sufficient in three dimensions, where molecules can approach one another along any axis. Nevertheless, the losses can still be suppressed by \emph{shielding} techniques, which engineer long-range repulsive barriers between molecules by using external fields to produce near-degeneracies between selected states of the molecular pair. This can be done with either static electric \cite{Avdeenkov:2006, Wang:dipolar:2015, Gonzalez-Martinez:adim:2017, Mukherjee:CaF:2023, Mukherjee:alkali:2024} or microwave \cite{Karman:shielding:2018, Lassabliere:2018, Karman:shielding-imp:2019, Karman:ellip:2020, Karman:res:2022, Deng:microwave:2023} fields.

Shielding techniques provide exquisite control over the long-range anisotropic dipole-dipole interaction between polar molecules. They have paved the way for the experimental realization of stable ultracold gases of polar molecules with high phase-space density \cite{Matsuda:2020, Li:KRb-shield-3D:2021, Schindewolf:NaK-degen:2022, Bigagli:NaCs:2023, Lin:NaRb:2023} and for the achievement of Fermi degeneracy \cite{Schindewolf:NaK-degen:2022} and Bose-Einstein condensation (BEC) \cite{Bigagli:BEC:NaCs:2024} for dipolar molecules in three dimensions. These advances have opened up a new playground to explore many-body physics with shielded polar molecules \cite{Jin:Jastrow:2024, Langen:dipolar-droplets:2025}.

Theories of many-body physics are often formulated in terms of zero-range contact potentials \cite{Huang:1957}, characterized by the s-wave scattering length $a$ \cite{Hutson:res:2007, Chin:RMP:2010}. For dipolar systems, this is supplemented by a long-range anisotropic dipole-dipole interaction \cite{Yi:2000, Yi:2001, Ronen:2006}, characterized by a dipole length  $a_\textrm{dd}$ \footnote{The dipole length usually used in mean-field theories \cite{Lahaye:2009, Chomaz:2023, Langen:dipolar-droplets:2025} is $a_\textrm{dd} = m d_\textrm{eff}^2/ (12 \pi \epsilon_{0} \hbar^2)$, where $m$ is the molecular mass. This is related to the dipole length $D$ that is common in scattering theory \cite{Bohn:BCT:2009, Mukherjee:CaF:2023} by $a_\textrm{dd} = 2D/3$.} associated with the effective dipole moment $d_\textrm{eff}$ \cite{Lahaye:2009, Chomaz:2023, Langen:dipolar-droplets:2025}. The interplay between $a$ and $a_\textrm{dd}$ is characterized by $\epsilon_\textrm{dd} = a_\textrm{dd}/a$, which determines deviations from the mean-field limit for dipolar quantum gases \cite{Lahaye:2009, Baranov:2012, Bottcher:2019, Chomaz:2023}. Shielded molecules offer much larger values of $\epsilon_\textrm{dd}$ than magnetic atoms \cite{Chomaz:2023}. They also offer the possibility of controlling $a$ and $a_\textrm{dd}$ independently, so provide a versatile platform for studying dipolar quantum matter \cite{Lahaye:2009, Baranov:2012, Chomaz:2023, Schmidt:2022}. However, the effective potentials for shielded molecules have much larger repulsive cores ($\gtrsim 10^3$~bohr) \cite{Deng:microwave:2023, Bigagli:NaCs:2023, Jin:Jastrow:2024, Langen:dipolar-droplets:2025} than for magnetic atoms. The resulting finite-range effects, together with quantum fluctuations, play a key role in stabilizing quantum phases of strongly interacting dipolar molecules \cite{Schmidt:2022, Jin:Jastrow:2024, Langen:dipolar-droplets:2025}.

Static-field shielding provides different degrees of shielding \cite{Gonzalez-Martinez:adim:2017} and different tunability of the scattering length \cite{Mukherjee:CaF:2023, Mukherjee:alkali:2024} for molecules with different dipole moments and rotational constants. It produces not only a long-range repulsion between molecules, but also a dipolar attraction at even longer range. For some molecules, the resulting potential well is strong enough to support one or more bound states \cite{Mukherjee:alkali:2024}, with associated poles in the scattering length at fields where the bound states cross threshold. Similar features occur in microwave shielding \cite{Lassabliere:2018, Quemener:electroassoc:2023}, and Chen \emph{et al.} \cite{Chen:field-linked-resonances:2023, Chen:field-linked-states:2024} have recently observed the bound states for fermionic Na$^{40}$K with collisions shielded using microwaves with elliptical polarization.

The purpose of the present paper is to point out that microwave shielding, in contrast to static-field shielding, is almost universal. With the appropriate choice of scaling for lengths and energies, it produces very similar loss rates, very similar effective potentials, and very similar tunability of the scattering length for all polar molecules. Deviations from universality arise only from couplings to well-separated rotational states that are not directly involved in shielding. We explain the origins of the universality and quantify the deviations from it for a range of molecules of current interest. The deviations are small when the Rabi frequencies used for shielding are much smaller than the molecular rotational constant.

The structure of this paper is as follows. Section \ref{sec:theory} describes the Hamiltonians and coupled-channel methods that we use to calculate scattering properties, including rate coefficients and scattering lengths. Section \ref{subsec:fully universal} demonstrates the universal behavior by comparing coupled-channel results for microwave-shielded NaCs and NaRb in a restricted basis set. Section \ref{subsec:deviations} then explores the small \emph{deviations} from universality that appear with larger basis sets. Finally, Sections \ref{subsec:fermions}, \ref{subsec:ellipt} and \ref{subsec:2field} consider extensions to handle identical fermions, elliptical polarization, and multiple polarizations. Section \ref{sec:conc} presents our conclusions and perspectives for future work.

\section{Theory}
\label{sec:theory}

\subsection{Coupled-channel formalism for two-body collisions in a microwave field }
\label{subsec:reduced_cc}

The theory of microwave shielding has been described in refs.\ \cite{Karman:shielding:2018, Karman:shielding-imp:2019}, and only a brief summary is given here to provide a foundation for the present work.

The total Hamiltonian that describes two-body interactions is
\begin{eqnarray}
\hat{H} = \frac{\hbar^2}{2\mu_\textrm{red}} \left ( - \frac{1}{R} \frac{d^{2}}{dR^{2}} R + \frac{\hat{\boldsymbol{L}}^2}{R^2} \right) + \hat{h}_\textrm{A} + \hat{h}_\textrm{B} + \hat{V}_\textrm{int}, \quad
\label{Eq: Dimer_Hamiltonian}
\end{eqnarray}
where $\mu_\textrm{red}$ is the reduced mass of the colliding pair, $R$ is the distance between the two molecules, and $\hat{\boldsymbol{L}}$ is the operator for relative angular momentum of the pair.
$\hat{h}_\textrm{A}$ and $\hat{h}_\textrm{B}$ are the internal Hamiltonians of the individual molecules in the presence of a microwave radiation field. The molecule-microwave interaction term for molecule $k$ is \citep{Cohen-Tannoudji:API:1998, Alyabyshev:microwave:2009}
\begin{eqnarray}
\hat{H}^{(k)}_\sigma &=&  - \frac{F_\textrm{ac}}{2 \sqrt{N_0}} [\hat{\mu}_{k,\sigma} \hat{a}_{\sigma} + \hat{\mu}_{k,\sigma}^\dagger \hat{a}_{\sigma}^\dagger] \nonumber \\
&+& \hbar \omega (\hat{a}_{\sigma} \hat{a}^{\dagger}_{\sigma}  - N_0),
\label{Eq: H_{AC}}
\end{eqnarray}
where $F_\textrm{ac}$ is the time-varying electric field and $N_0$ is the reference number of photons. Here, $\hat{a}^{\dagger}_{\sigma}$ and $\hat{a}_{\sigma}$ are the creation and annihilation operators for photons with polarization $\sigma$ and $\hat{\mu}_{k,\sigma}^\dagger$ and $\hat{\mu}_{k,\sigma}$ are the corresponding components of the dipole moment operator. Here we focus on shielding with circular polarization, $\sigma^+$. The frequency $\omega$ is blue-detuned from the rotational transition $n=0 \rightarrow 1$ by $\Delta$. The interaction strength is characterized by the Rabi frequency $\Omega = F_\textrm{ac} \mu/(\sqrt{3}\hbar)$, where $\mu$ is the body-fixed dipole moment of a single molecule. The resulting dressed-monomer states $\vert\textrm{+}\rangle$ and $\vert\textrm{--}\rangle$ are separated by energy $\hbar [\Omega^{2}+\Delta^{2}]^{1/2}$. The ratio $\Delta/\Omega$ serves as a control knob to tune the effective dipole moment $d_\textrm{eff}=\mu /\left[12 \left(1+(\Delta/\Omega)^2\right)\right]^{1/2}$ that governs the overall strength of the dipole-dipole interaction between shielded molecules \cite{Karman:res:2022}.

The interactions that are important for shielding occur principally in the long-range region, at $R \gg 100$~bohr. We therefore approximate $V_\textrm{int}$ by the dipole-dipole interaction,
\begin{eqnarray}
\hat{H}_\textrm{dd} = -\frac{\sqrt{6}}{4 \pi \epsilon_0 R^3} T^{(2)}(\hat{\mu}_\textrm{A},\hat{\mu}_\textrm{B})
\cdot C^{(2)}(\boldsymbol{\hat{R}}),
\label{Eq: Dipole-Dipole Operator}
\end{eqnarray}
where $T^{(2)}$ and $C^{(2)}$ are second-rank tensors, $\boldsymbol{\hat{R}}$ is a unit vector along the intermolecular axis and the components of $C^{(2)}(\boldsymbol{\hat{R}})$ are Racah-normalized spherical harmonics. Other terms in $V_\textrm{int}$, such as those involving higher-order multipole moments and electronic dispersion interactions, die off faster with $R$ and make only small contributions at the distances important for  shielding.

The total wavefunction is expanded as
\begin{eqnarray}
\Psi (R, \boldsymbol{\hat{R}}, \boldsymbol{\hat{r}}_\textrm{A}, \boldsymbol{\hat{r}}_\textrm{B}) = \frac{1}{R} \sum_{j} \psi_j (R) \Phi_{j} (\boldsymbol{\hat{R}}, \boldsymbol{\hat{r}}_\textrm{A}, \boldsymbol{\hat{r}}_\textrm{B}),
\label{Eq: Total Wavefunction}
\end{eqnarray}
where $\boldsymbol{\hat{r}}_k$ is the unit vector along the axis of molecule $k$. In the uncoupled representation, the channel basis functions $\Phi_j$ are products of rotational pair-state functions, photon number states $N$ (counted relative to $N_0$) and eigenfunctions of $\hat{\boldsymbol{L}}^2$. They are symmetrized for exchange of identical particles as necessary. It should be noted that a basis set constructed from field-dressed pair functions, as here, is \emph{not} always the same as one constructed from pairs of field-dressed single-molecule functions. There is in reality a single photon bath that is shared between the molecules, which cannot in general be decomposed into separate baths for the two molecules.

Substituting the expansion (\ref{Eq: Total Wavefunction}) into the total Schr\"odinger equation yields the coupled-channel equations, which are a set of coupled differential equations in $R$ that we solve as described below.  For $\sigma^+$ polarization in the $xy$ plane, the projection of the total angular momentum $M_\textrm{tot}=m_{n_\textrm{A}}+m_{n_\textrm{B}}+M_L+N$ onto $z$ is a conserved quantity. We therefore carry out calculations for a single value of $M_\textrm{tot}$ at a time.

In the present work, basis functions for end-over-end angular momentum are included up to $L_\textrm{max}=12$. This gives convergence of rate coefficients within 1\%. The values of $M_\textrm{tot}$ required for convergence vary with collision energy; at the highest energies considered, we include $M_\textrm{tot}$ from $-8$ to $+8$. We neglect spin degrees of freedom, whose effects can usually be suppressed by adding a magnetic field $B \gtrsim 100$ G \cite{Karman:shielding:2018}.

\begin{figure}[tb]
\includegraphics[width=\columnwidth]{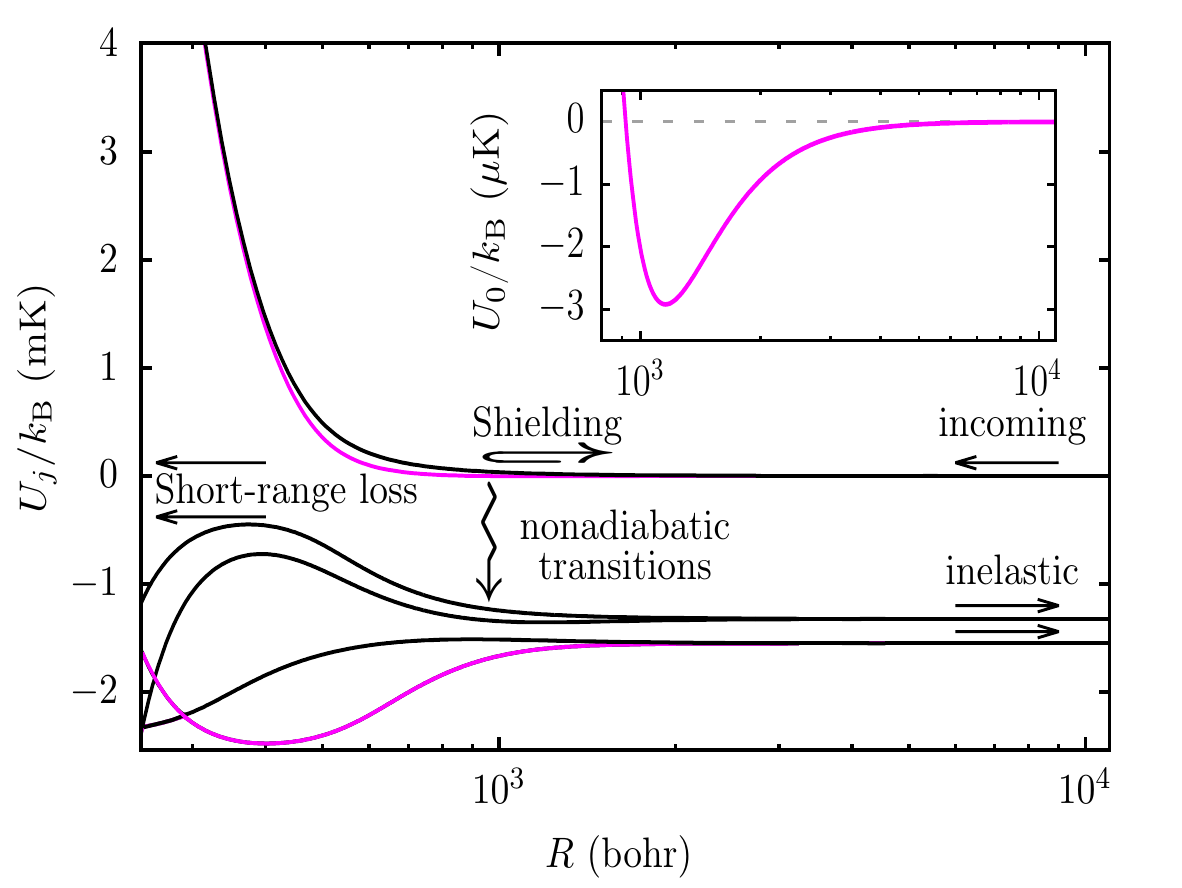}
\caption{Adiabatic curves $U_j(R)$ important for microwave shielding, showing the mechanisms involved in shielding and 2-body loss. The curves shown are for NaRb with $\Omega=\Delta=22.8$~MHz, but are qualitatively similar for other molecules (with molecule-dependent scales for length and energy). Magenta lines indicate s-wave channels. The inset shows an expanded view of the potential minimum in the incoming s-wave adiabat.}
\label{fig:schematic-adiabats}
\end{figure}

\begin{table*}[tb]
\caption{Scaling and other parameters for the molecules considered in this study. The body-fixed dipole moments and rotational constants are taken from refs.\ \cite{Dagdigian:1972, Docenko:2006, Guo:NaRb:2016, Guo:NaRb:2018, Russier-Antoine:2000, Zuchowski:vdW:2013}.
\label{tab:scaling_param}} \centering
\begin{ruledtabular}
\begin{tabular}{cccccc}
& $R_3$ (bohr)  & $E_3/k_\textrm{B}$ (K) & $k_3$ (cm$^3$ s$^{-1}$) & $\tilde{b}$ & $\mu$ (D)
 \\ \hline
$^{23}$Na$^{133}$Cs &   $9.9 \times 10^5$ & $1.1 \times 10^{-12}$ & $5.4 \times 10^{-7}$ & $7.4 \times 10^{10}$ & 4.75 \\
$^{23}$Na$^{87}$Rb  &   $3.2 \times 10^5$ & $1.6 \times 10^{-11}$ & $2.4 \times 10^{-7}$ & $6.4 \times 10^9$ & 3.2  \\
$^{23}$Na$^{39}$K   &   $1.3 \times 10^5$ & $1.7 \times 10^{-10}$ & $1.8 \times 10^{-7}$ & $8.2 \times 10^8$ & 2.72  \\
$^{6}$Li$^{133}$Cs  &   $1.1 \times 10^6$ & $9.6 \times 10^{-13}$ & $6.9 \times 10^{-7}$ & $3.3 \times 10^{11}$ & 5.39  \\
\end{tabular}
\end{ruledtabular}
\end{table*}

\subsection{Origin of universality}
\label{subsec:origin}

The coupled-channel equations take a particularly simple form when expressed in terms of dimensionless reduced lengths $\tilde{R}=R/R_3$, reduced energies $\tilde{E}=E/E_3$, reduced Rabi frequencies $\tilde{\Omega}=\hbar\Omega/E_3$ and detunings $\tilde{\Delta}=\hbar\Delta/E_3$. Here $R_3$ and $E_3$ are scaling factors introduced by Gonz\'alez-Mart\'{\i}nez \emph{et al.}\ \cite{Gonzalez-Martinez:adim:2017} in the context of static-field shielding, $R_3=(2 \mu_\textrm{red}/ \hbar^2)(\mu^2/4 \pi \epsilon_0)$ and $E_3=\hbar^2/(2 \mu_\textrm{red} R_3^2)$.

The physics of microwave shielding is dominated by channels arising from field-dressed pair states corresponding to $(n_\textrm{A},n_\textrm{B},N)$= (0,0,0), (1,0,$-1$) and (1,1,$-2$). For identical molecules there are 10 such pair states \cite{Deng:microwave:2023}, which are brought near to degeneracy by the microwave field. Channels arising from other values of $(n_\textrm{A},n_\textrm{B},N)$ are much further away in energy and much less important: provided $\Omega, \Delta\ll b_\textrm{rot}/\hbar$, they contribute only weakly to shielding. In the dimensionless representation described above, the coupled-channel equations involving this near-resonant set of 10 pair states are completely independent of the molecular properties. This is the origin of universality for microwave shielding.

The universality of microwave shielding contrasts with the nonuniversality of static-field shielding \cite{Gonzalez-Martinez:adim:2017, Mukherjee:alkali:2024}. Static-field shielding relies on creating near-degeneracies between field-dressed pair states that are principally $(n_\textrm{A},m_{n_\textrm{A}})$+$(n_\textrm{B},m_{n_\textrm{B}})$ = (1,0)+(1,0) and (0,0)+(2,0). The electric field needed to achieve the near-degeneracy is approximately $F_\textrm{X}=3.24 b_\textrm{rot}/\mu$. At a particular value of $\tilde{F}=F\mu/b_\textrm{rot}$, \emph{all} the separations between asymptotic channels are proportional to $b_\textrm{rot}$, or in reduced units to $\tilde{b}=b_\textrm{rot}/E_3$. This includes the separation between (1,0)+(1,0) and (0,0)+(2,0). Because of this, key features of the shielding potential depend strongly on $\tilde{b}$: these include the position of the repulsive wall, the depth of the potential well, and the height of the shielding barrier \cite{Mukherjee:alkali:2024}. As a result, static-field shielding depends strongly on $\tilde{b}$: it is effective when $\tilde{b}$ is large, and ineffective when it is small ($\tilde{b}\lesssim 10^7$) \cite{Gonzalez-Martinez:adim:2017}. Moreover, static-field-shielded molecules with $\tilde{b}\gtrsim 10^{10}$ can support two-molecule bound states at long range \cite{Mukherjee:alkali:2024}, while those with smaller $\tilde{b}$ cannot.

\subsection{Scattering lengths, cross sections, and rate coefficients}

Shielding techniques can often suppress collisional losses enough that elastic collision rates exceed the two-body loss rates by orders of magnitude. The overall efficiency of shielding can be parametrized by the ratio $\gamma$ of elastic and total loss rates. The loss may occur in three main ways. First, one or both molecules may be transferred to lower internal states; such inelastic processes release kinetic energy, and both molecules are usually ejected from the trap. Secondly, colliding pairs may reach short range ($R\lesssim 100$~bohr), where they are likely to be lost as described in the Introduction. Lastly, molecules may be lost through 3-body collisions, which may produce either 3-body recombination or additional inelasticity.

The mechanisms of elastic scattering and loss are best understood in terms of the effective potentials (adiabats) for the different channels involved. The adiabats $U_j(R)$ are defined as the eigenvalues of $\hat{h}_\textrm{A} + \hat{h}_\textrm{B} + \hat{V}_\textrm{int} +\hbar^2\hat{\boldsymbol{L}}^2/(2\mu_\textrm{red}R^2)$ as functions of $R$, and examples of the key ones for microwave shielding are shown in Fig.\ \ref{fig:schematic-adiabats}. In the presence of shielding, the adiabats for the incoming channels exhibit repulsive walls or potential barriers at moderately long range. Colliding pairs that remain on the incoming adiabats are usually reflected back to long range (elastic scattering), but may also tunnel through the barrier to reach short range. If they reach short range, they are likely to be lost as described in the Introduction. Alternatively, colliding pairs may undergo nonadiabatic transitions to lower pair states. In this case, they may again reach short range (and contribute to short-range loss) or be reflected back to long range in the lower channel (causing inelastic loss). These nonadiabatic processes are usually the dominant source of loss for microwave shielding.

To model these processes, we solve the coupled-channel equations subject to a short-range boundary condition that absorbs all flux that reaches short range \cite{Clary:1987, Janssen:PhD:2012, Mukherjee:CaF:2023}. We have implemented the coupled equations for microwave shielding as a plug-in basis-set suite for the MOLSCAT package \cite{molscat:2019, mbf-github:2023}. We solve them using log-derivative propagators \cite{Manolopoulos:1986, Alexander:1987} adapted to co-propagate two linearly independent solutions for each channel, and use these to construct traveling-wave solutions with no outgoing part at a distance $R_\textrm{absorb} = 5 \times 10^{-5}\,R_3$. This produces a nonunitary S matrix $\boldsymbol{S}$ for each value of the microwave parameters ($\tilde{\Omega}$ and $\tilde{\Delta}/\tilde{\Omega} = \Delta/\Omega$) and the reduced collision energy $\tilde{E}_\textrm{coll}=E_\textrm{coll}/E_3$. Cross sections for elastic scattering ($\sigma_\textrm{el}$), inelastic loss ($\sigma_\textrm{inel}$) and short-range loss ($\sigma_\textrm{sr}$) are obtained from the S-matrix elements as described in Ref.\ \cite{Mukherjee:CaF:2023}. The sum of $\sigma_\textrm{inel}$ and $\sigma_\textrm{sr}$ is $\sigma_\textrm{tot}$ and represents the total two-body loss. The corresponding rate coefficients are $k=v\sigma$, where $v=(2E_\textrm{coll}/\mu_\textrm{red})^{1/2}$. We also calculate complex energy-dependent s-wave scattering lengths \citep{Hutson:res:2007},
\begin{eqnarray}
a(k_0) = \alpha(k_0)-i\beta(k_0) = \frac{1}{i k_0} \left( \dfrac{1-S_{00} (k_0)}{1+S_{00} (k_0)} \right),
\end{eqnarray}
where $k_0 = (2\mu_\textrm{red}E_\textrm{coll}/\hbar^2)^{1/2}$ is the incoming wavevector. We define the corresponding reduced collision properties as $\tilde{\alpha} = \alpha/R_3$, $\tilde{\beta} = \beta/R_3$, $\tilde{\sigma}=\sigma/(4\pi R_3^2)$, and $\tilde{k}= k/k_3$, where $k_3= 4 \pi R_3 \hbar/\mu_\textrm{red}$.

It should be noted that the adiabats we consider here are functions only of $R$. This differs from the treatment of Deng \emph{et al.}\ \cite{Deng:microwave:2023}, who derive adiabatic \emph{surfaces} that are explicit functions of $R$, $\theta$, and $\phi$, where $\theta$ and $\phi$ are spherical polar angles that describe the orientation of $\hat{\boldsymbol{R}}$. Moreover, Deng \emph{et al.}\ perform most of their calculations on a single adiabatic surface and do not consider loss processes.

\section{Results}
\label{sec:results}

In this section, we explore both the universal behavior of microwave shielding and deviations from it for a selection of ultracold polar molecules of current experimental interest. We choose the molecules NaK, NaRb and NaCs, which have a wide range of dipole moments and rotational constants. Their scaling parameters and values of $\tilde{b}$ are summarized in Table \ref{tab:scaling_param}.

\begin{figure*}[tb]
\includegraphics[width=2.1\columnwidth]{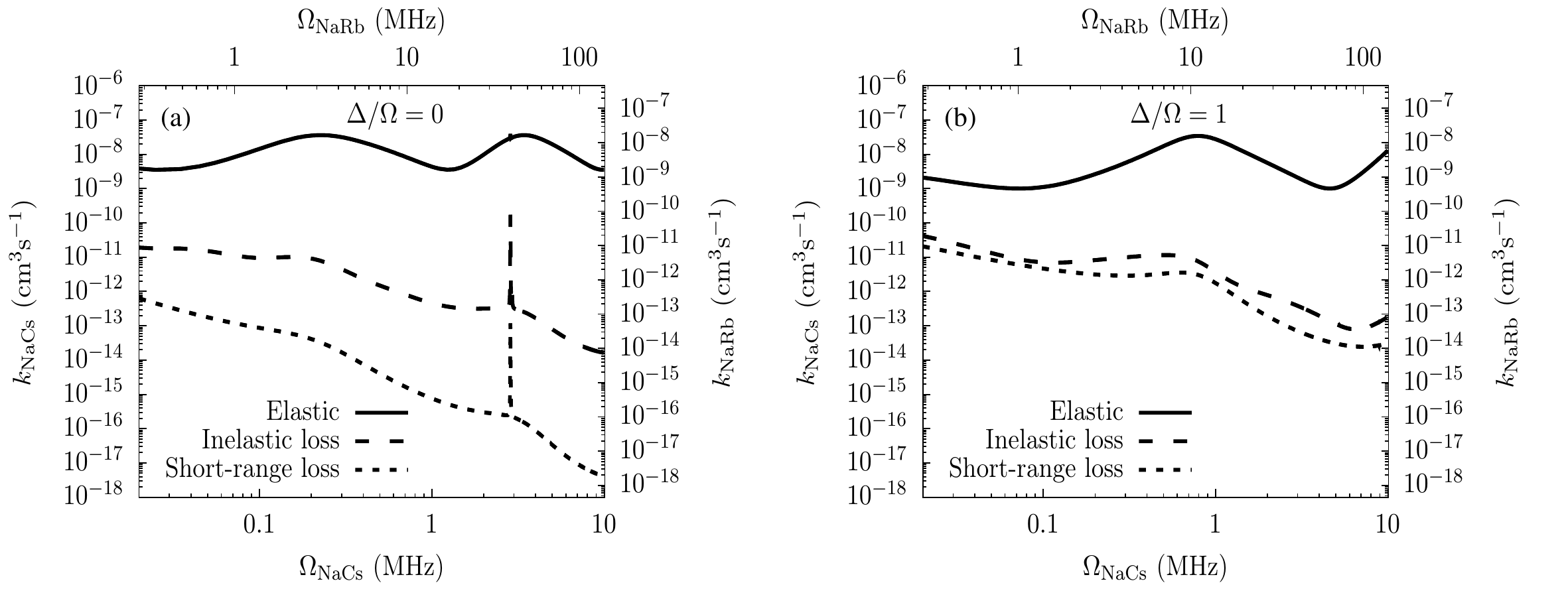}
\caption{Rate coefficients for NaCs and NaRb as a function of $\Omega$ at $E_\textrm{coll}=1000 E_3$ for (a) $\Delta/\Omega=0$; (b) $\Delta/\Omega=1$, obtained with the 10-pair basis set and demonstrating universal behavior.}
\label{fig:univ-rate-Rabi}
\end{figure*}

\begin{figure*}[tb]
\centering
\includegraphics[width=2.1\columnwidth]{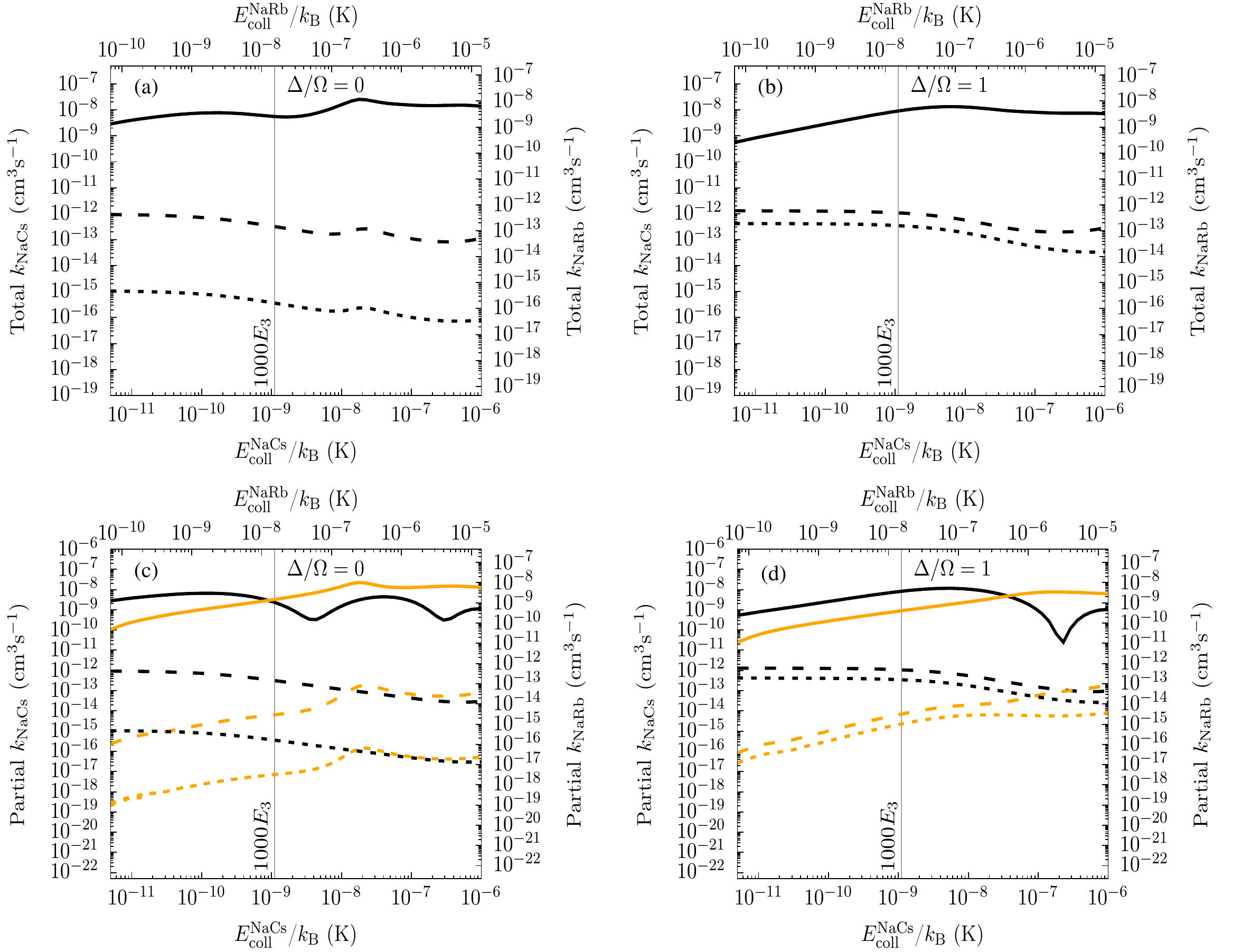}
\caption{Rate coefficients as a function of collision energy at $\tilde{\Omega}=7\times10^7$, corresponding to $\Omega=1.65$ MHz for NaCs and 22.8 MHz for NaRb, obtained with the 10-pair basis set and demonstrating universal behavior. The upper panels show rate coefficients summed over $L_\textrm{in}$ for (a) $\Delta/\Omega=0$ and (b) $\Delta/\Omega=1$. Elastic, inelastic, and short-range loss are represented by solid, dashed and dotted lines, respectively. The lower panels (c) and (d) show the contributions from $L_\textrm{in}=0$ (black lines) and $L_\textrm{in}>0$ (orange lines) separately. The vertical lines indicate the collision energy $E_\textrm{coll}=1000E_3$ used in Fig.\ \ref{fig:univ-rate-Rabi}.}
\label{fig:univ-rate-energy}
\end{figure*}

\subsection{Fully universal regime}
\label{subsec:fully universal}

To demonstrate fully universal behavior, we use a basis set that includes only the 10 pair states described in Section \ref{subsec:origin}, formed from symmetrized combinations of $(n_\textrm{A},n_\textrm{B},N) = (0,0,0)$, (1,0,$-1$) and (1,1,$-2$). The monomer states $(n,m_n)=(1,0)$ and $(1,-1)$ are not coupled to (0,0) by a microwave field with $\sigma^+$ polarization; they are referred to as the dark states. The 3 pair states involving two dark states are not coupled to the remainder by $\hat{H}_\textrm{dd}$, and could be omitted from the calculation for a single microwave field; however, we include them here to maintain generality for future extensions.

\subsubsection{Rate coefficients}

Figure \ref{fig:univ-rate-Rabi} shows rate coefficients for elastic scattering, inelastic loss and short-range loss for microwave-shielded NaCs and NaRb as functions of Rabi frequency. The two panels show cross sections for $\Delta/\Omega=0$ and 1; the former produces the largest values of $d_\textrm{eff}$, while the latter is typical of recent experiments \cite{Lin:NaRb:2023, Bigagli:NaCs:2023}. The calculations are at reduced collision energy $E_\textrm{coll}=1000E_3$, corresponding to $E_\textrm{coll}\approx1.1$ nK for NaCs and 16 nK for NaRb. Universality applies to both s-wave collisions ($L=0$) and to collisions with higher partial waves, $L>0$; this energy is chosen so that both $L=0$ and 2 make significant contributions to the elastic cross sections for $\Delta/\Omega=0$.

The value of $E_3$ for NaRb is about a factor of 15 larger than for NaCs, while $k_3$ is about a factor of 2.1 smaller. Accordingly, the axes for Rabi frequencies and rate coefficients (which are logarithmic in Fig.\ \ref{fig:univ-rate-Rabi}) are shifted by these factors between the two molecules. In this representation, the rate coefficients appear identical for the two molecules but are read off different axes for each molecule. This demonstrates full universality with the 10-pair basis set.

For $\Delta/\Omega = 0(1)$, the ratio $\gamma$ between the rates for elastic scattering and total loss reaches $5\times10^3$ for $\tilde{\Omega}\approx1\times10^7$ $(4\times10^7)$, corresponding to 0.25 (0.95) MHz for NaCs and 3.46 (13.2) MHz for NaRb. These values are readily extended to any other polar molecule.

The narrow peaks that appear near $\Omega=2.87$ MHz for NaCs and 39.7 MHz for NaRb at $\Delta/\Omega=0$ arise from a d-wave shape resonance in the incoming channel. It is noteworthy that even this feature appears at exactly the same reduced frequency and with exactly the same amplitude for both molecules.

Figure \ref{fig:univ-rate-energy} demonstrates universality as a function of collision energy, in both s-wave and higher-partial-wave regimes. Here we choose $\tilde{\Omega}=7\times 10^7$, corresponding to $\Omega=1.65$ MHz for NaCs and 22.8 MHz for NaRb. The upper panels show the rate coefficients summed over partial waves, while the lower panels show the breakdown into s-wave and higher-$L$ components. The axes for the two molecules are again shifted so that each molecule is represented by the same lines read off different axes.

\begin{figure}[tb]
\includegraphics[width=\columnwidth]{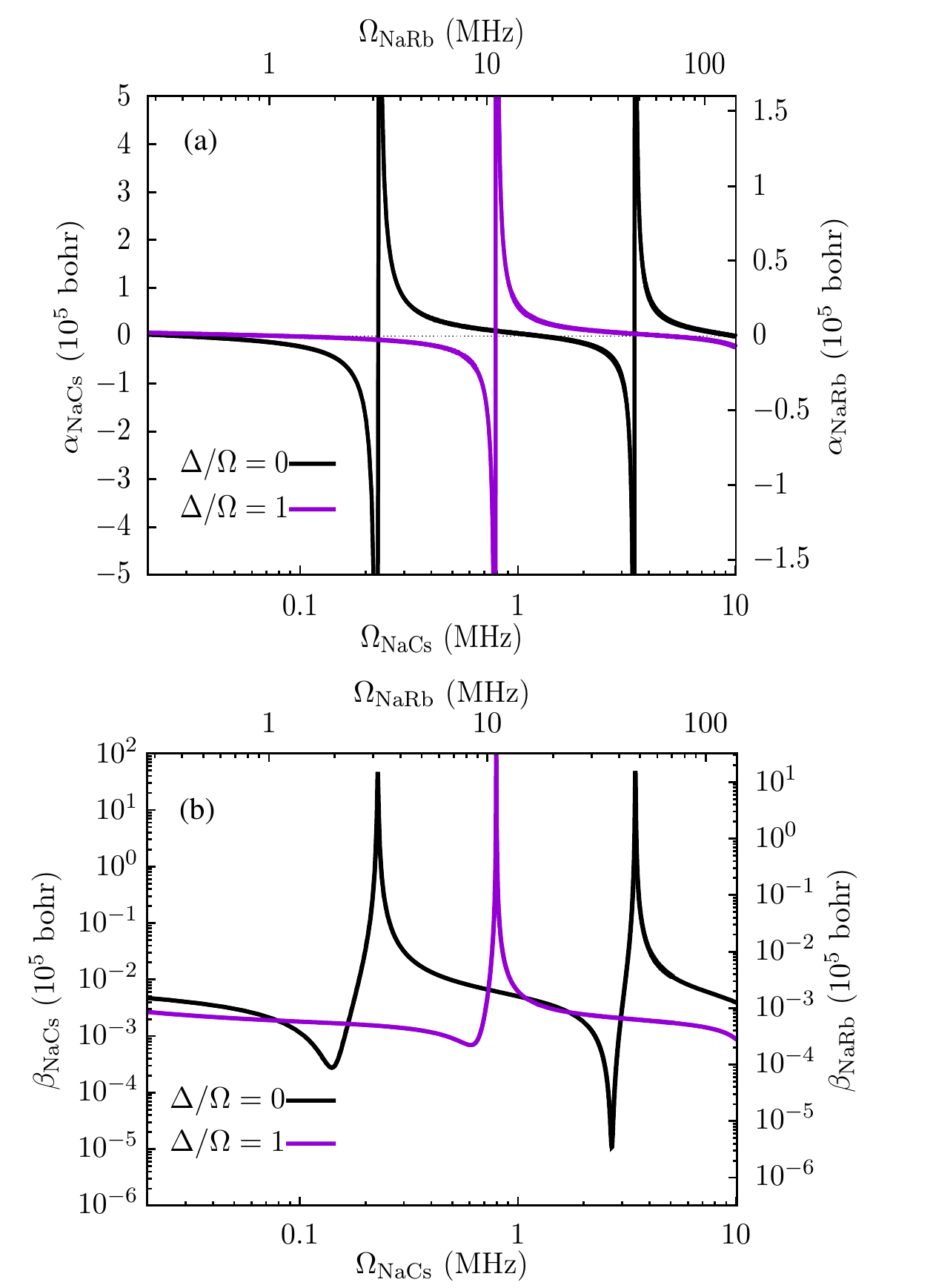}
\caption{(a) Real ($\alpha$) and (b) imaginary ($\beta$) parts of the scattering length for NaCs and NaRb as a function of $\Omega$ for $\Delta/\Omega=0$ and 1 at $E_\textrm{coll}=1000 E_3$, obtained with the 10-pair basis set and demonstrating universal behavior.}
\label{fig:univ-scatlen}
\end{figure}

\subsubsection{Scattering lengths and effective potentials}

Figure \ref{fig:univ-scatlen} shows the real and imaginary parts of the s-wave scattering lengths for NaCs and NaRb as functions of Rabi frequency, calculated at $E_\textrm{coll}=1000E_3$ with the 10-pair basis set. Once again the results are identical for the two molecules with appropriately scaled axes, demonstrating full universality with this basis set.

The universality of the scattering length is a manifestation of a deeper universality in the effective potentials (adiabats) for shielded molecules. In the 10-pair basis set, the entire set of coupled equations is universal, so the adiabats are as well. Figure \ref{fig:univ-adiabat} shows the adiabats $U_0(R)$ for s-wave scattering at the incoming threshold, for $\Delta/\Omega=0$ and 1 and the same Rabi frequencies as Fig.\ \ref{fig:univ-rate-energy}. The entire curves are universal, including features such as well depths and turning points.

\begin{figure}[tb]
\includegraphics[width=\columnwidth]{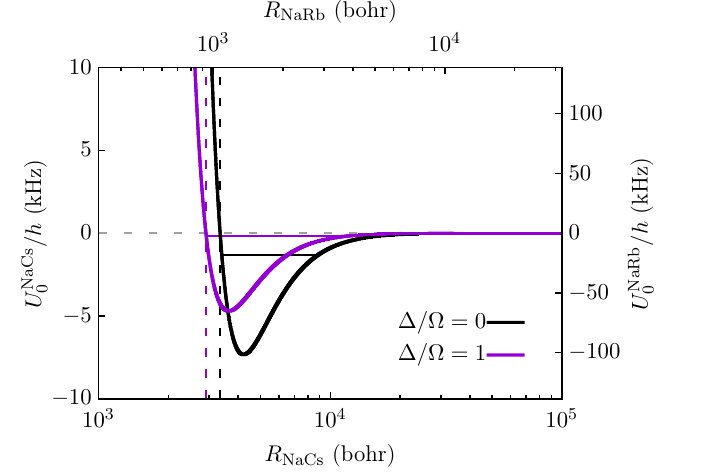}
\caption{Shielding adiabats $U_0 (R)$ for $\tilde{\Omega} = 7\times10^7$, corresponding to $\Omega=1.65$ MHz for NaCs and 22.8 MHz for NaRb, obtained with the 10-pair basis set and demonstrating universal behavior for $\Delta/\Omega=0$ (black) and 1 (violet). The dashed vertical lines show the inner turning points $R_\textrm{t}$ at zero collision energy. The solid horizontal lines show the energies of 2-molecule bound states.}
\label{fig:univ-adiabat}
\end{figure}

An important feature of Fig.\ \ref{fig:univ-scatlen}(a) is that the poles and zeroes of $\alpha$ occur at quite different Rabi frequencies for $\Delta/\Omega=0$ and 1. In a single-channel picture where the collision takes place entirely on $U_0(R)$, the zero-energy scattering length $a$ may be approximated semiclassically in terms of a phase integral \cite{Mukherjee:SU(N):2025},
\begin{equation}
a = R_\textrm{t} -
\sqrt{\frac{8}{15}} D
\tan\left(\Phi-\frac{\pi}{4}\right)
\label{eq:aPhi}
\end{equation}
where $R_\textrm{t}$ is the inner turning point at zero energy, $D=(3/2)a_\textrm{dd}=\mu_\textrm{red}d_\textrm{eff}^2/(4\pi\epsilon_0\hbar^2)$,
and the phase integral is
\begin{equation}
\Phi = \int_{R_\textrm{t}}^\infty k_0(R) dR,
\label{eq:Phi}
\end{equation}
where $\hbar^2k_0(R)^2/(2\mu_\textrm{red}) = -U_0(R)$. Poles in $a$ thus appear when $\Phi/\pi$ passes through $n+3/4$ for integer $n$, and small values ($a\approx R_\textrm{t}$) appear when it passes through $n+1/4$. In multichannel scattering for well-shielded molecules, these equations apply approximately to the real part of the scattering length. However, the single-channel approximation starts to breaks down when losses are substantial, as then the resonances are decayed \cite{Hutson:res:2007} and the scattering length is complex, with finite-amplitude oscillations in $\alpha$ and $\beta$ rather than poles.

\begin{figure}[tb]
\includegraphics[width=\columnwidth]{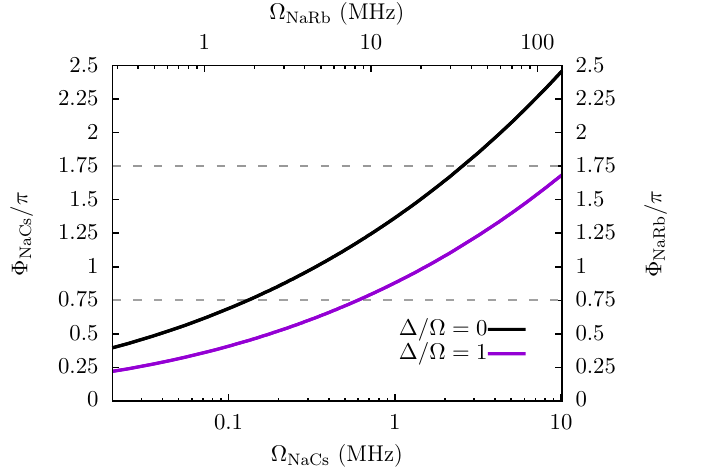}
\caption{Phase integral $\Phi$ for shielding adiabats of NaCs and NaRb as a function of $\Omega$ at $\Delta/\Omega=0$ (black) and $\Delta/\Omega=1$ (violet), obtained with the 10-pair basis set and demonstrating universal behavior. The horizontal dashed lines indicate the values of $\Phi$ where the first and second bound states and the corresponding zero-energy resonances appear in a semiclassical approximation.}
\label{fig:univ-Phi}
\end{figure}

\begin{figure}[tb]
\includegraphics[width=\columnwidth]{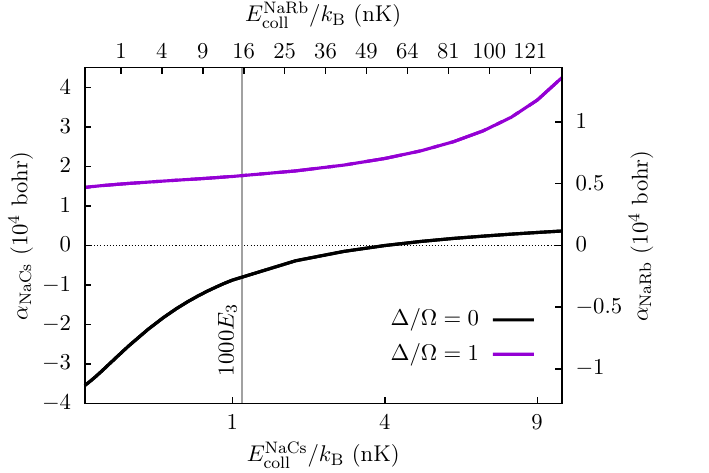}
\caption{Real part of scattering length $\alpha$ as a function of $E_\textrm{coll}$ at  $\tilde{\Omega} = 7\times10^7$ corresponding to $\Omega=1.65$ MHz for NaCs and 22.8 MHz for NaRb, obtained with the 10-pair basis set and demonstrating universal behavior. }
\label{fig:univ-scatlen-E}
\end{figure}

Figure \ref{fig:univ-Phi} shows the phase integral $\Phi$ as a function of Rabi frequency for $\Delta/\Omega=0$ and 1, obtained from Eq.\ \ref{eq:Phi} using adiabats calculated with the 10-pair basis set. Once again the behavior is fully universal. The points at which $\Phi/\pi$ passes through $n+3/4$ and $n+1/4$ qualitatively explain the positions of the poles and zeroes in $\alpha$ in Fig.\ \ref{fig:univ-scatlen}. However, there are some shifts, both because the semiclassical expression (\ref{eq:aPhi}) is approximate and because Fig.\ \ref{fig:univ-scatlen} was obtained with coupled-channel calculations at $E_\textrm{coll}=1000E_3$, where the poles and zeroes are at somewhat higher Rabi frequency than at $E_\textrm{coll}=0$. Figure \ref{fig:univ-scatlen-E} shows the energy-dependence of $\alpha$ for $\tilde{\Omega}=7\times10^7$ for NaCs and NaRb with $\Delta/\Omega=0$ and 1, showing that it is once again universal. In the threshold regime, $\alpha$ is linear in $\tilde{E}_\textrm{coll}^{1/2}$; however, the gradient is lower and the near-linear regime extends to higher $\tilde{E}$ for $\Delta/\Omega=1$, because $d_\textrm{eff}$ is a factor of $\sqrt{2}$ smaller than for $\Delta/\Omega=0$.

Figures \ref{fig:univ-adiabat} and \ref{fig:univ-Phi} also demonstrate that the bound states supported by the long-range well are universal and will appear at the same values of $\tilde{\Omega}$ and $\Delta/\Omega$ for all molecules. The only difference between different systems is in the scaling factor $E_3$ and thus in the absolute values of $\Omega$ required for different molecules.

We have described universality in terms of the coupled equations for scattering and the resulting adiabatic curves $U_j(R)$. However, it also applies to the anisotropic adiabatic surfaces introduced by Deng \emph{et al.}\ \cite{Deng:microwave:2023} (their ``effective potentials"), which are functions of $\theta$ and $\phi$ as well as $R$. These 3-dimensional surfaces have been used for both fermions \cite{Deng:microwave:2023} and bosons \cite{Langen:dipolar-droplets:2025}. They will be very similar for different polar molecules when expressed in terms of reduced lengths $R/R_3$, reduced energies $E/E_3$ and reduced Rabi frequencies $\hbar\Omega/E_3$.

\subsection{Deviations from universality}
\label{subsec:deviations}

In this section, we explore deviations from universality for a selection of ultracold molecules of current experimental interest. To do this, we consider two different extended basis sets. The first of these is that described in Ref.\ \cite{Karman:shielding:2018}, with all rotational functions up to $n_\textrm{max}=1$ and photon numbers $N$ = 0, $-1$, $-2$; this contains 30 pair states, both above and below the near-resonant set. The second includes all rotational functions up to $n_\textrm{max}=2$; this contains 135 pair states, with the additional ones all lying above the near-resonant set.

\begin{figure*}[tb]
\includegraphics[width=2.1\columnwidth]{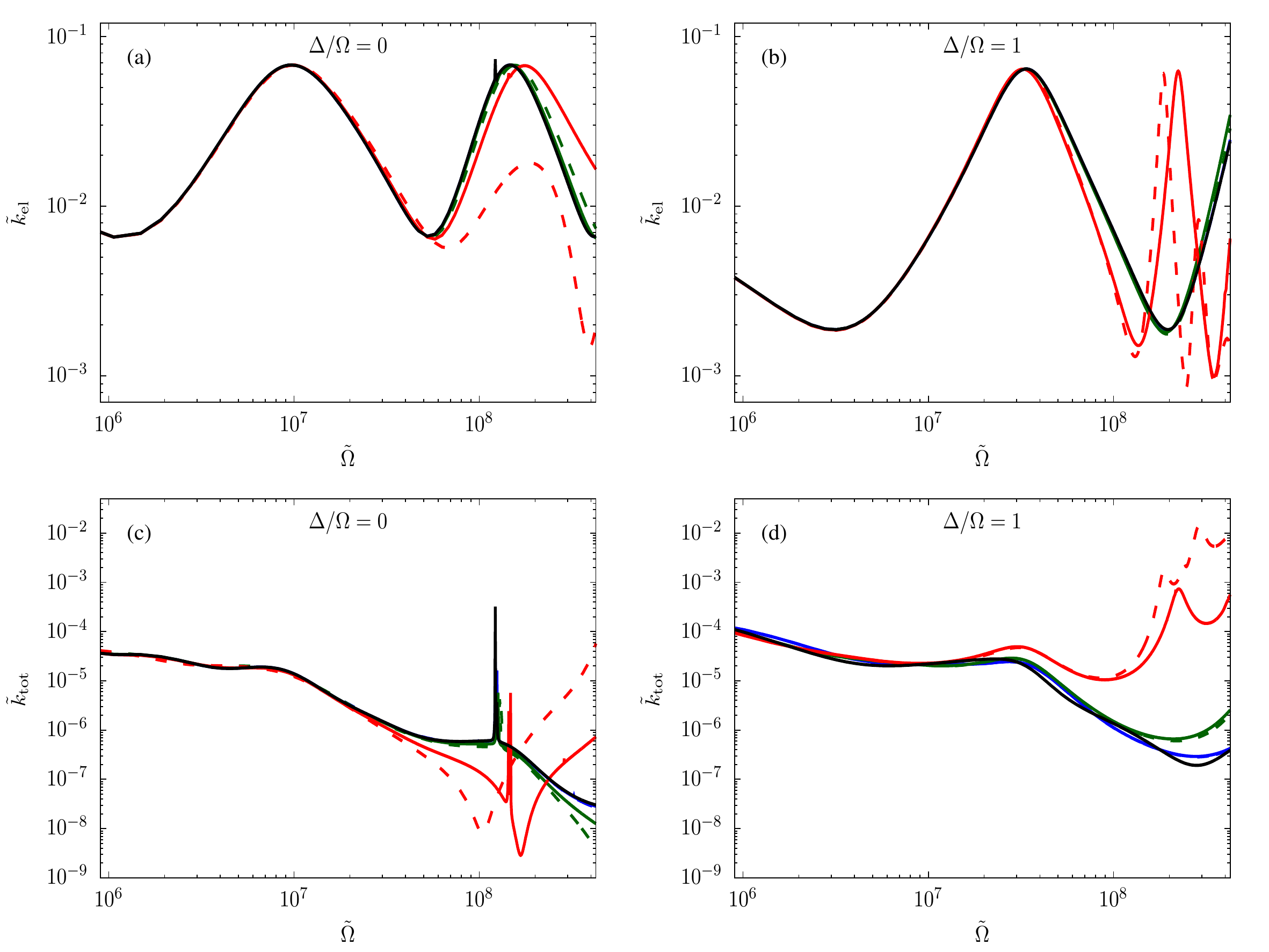}
\caption{Upper panels: Reduced rate coefficients $\tilde{k}_\textrm{el}$ for elastic scattering at $\tilde{E}_\textrm{coll}=1000$ for NaCs (blue), NaRb (green) and NaK (red), compared to the universal result (black) at (a) $\Delta/\Omega=0$ and (b) $\Delta/\Omega=1$. Lower panels: The corresponding reduced rate coefficients $\tilde{k}_\textrm{tot}$ for total loss. Solid and dashed lines show results obtained with extended basis sets with $n_\textrm{max}=1$ and $n_\textrm{max}=2$, respectively. In some cases the lines for individual molecules lie underneath the universal result.}
\label{fig:nonuniv-rates}
\end{figure*}

\subsubsection{Rate coefficients}

Figure \ref{fig:nonuniv-rates} shows rate coefficients for elastic scattering and total loss as a function of reduced Rabi frequency $\tilde{\Omega}$ for NaCs (blue), NaRb (green) and NaK (red), calculated with the two extended basis sets described above. These are compared with the fully universal result (black). All the results shown are for reduced collision energy $\tilde{E}_\textrm{coll}=1000$. It is evident that all the systems show nearly universal behavior at low $\tilde{\Omega}$, but start to deviate at high $\tilde{\Omega}$. The deviations are somewhat larger in loss rates than in elastic scattering, but generally similar in magnitude for $\Delta/\Omega=0$ and 1. They are significantly larger for the basis set with $n_\textrm{max}=2$ than for that with $n_\textrm{max}=1$.

The key molecular property that governs the deviations is the value of $\tilde{b}$, given in Table \ref{tab:scaling_param}. In reduced units, the 10 pair states that dominate microwave shielding are separated by energies that depend on $\tilde{\Omega}$  and $\tilde{\Delta}$ but are independent of the rotational constant $b$. However, the additional pair states that are responsible for deviations from universality are asymptotically separated from these 10 by energies proportional to $b$, or in reduced units to $\tilde{b}$. The main couplings between pair states are proportional to $\tilde{\Omega}$, so universality is generally accurate when $\tilde{\Omega}\ll\tilde{b}$. Figure \ref{fig:nonuniv-rates} shows that large deviations occur only when $\tilde{\Omega}$ is comparable to $\tilde{b}$ and that universality is accurate to within 5\% when $\tilde{\Omega}/\tilde{b}\lesssim 0.02$. For NaK, with $\tilde{b}=8.2\times 10^8$, universality starts to break down about half-way across Fig.\ \ref{fig:nonuniv-rates}. For NaCs, with $\tilde{b}=7.4\times 10^{10}$, the near-universal region extends almost to the right-hand edge of the Figure.

Some adiabats that correlate with rotationally excited pair states do come down far enough in energy to cross the principal ones at very short range, $\tilde{R}\lesssim 2 \times 10^{-3}$ for NaK and even smaller for molecules with larger $\tilde{b}$. This does not have much effect on the rate coefficients for elastic scattering or total loss, but it can change the details of the short-range physics. In particular, colliding pairs that undergo a nonadiabatic transition from the incoming channel to a lower one may penetrate to short range in one basis set (so contribute to short-range loss) but be reflected back to long range in another basis set (so contribute to inelastic loss). Nevertheless, both outcomes contribute to total loss. This is the reason that Fig.\ \ref{fig:nonuniv-rates} shows total loss, which is close to universal, instead of breaking it down into short-range and inelastic loss, which are individually more variable.

\begin{figure*}[tb]
\includegraphics[width=2.1\columnwidth]{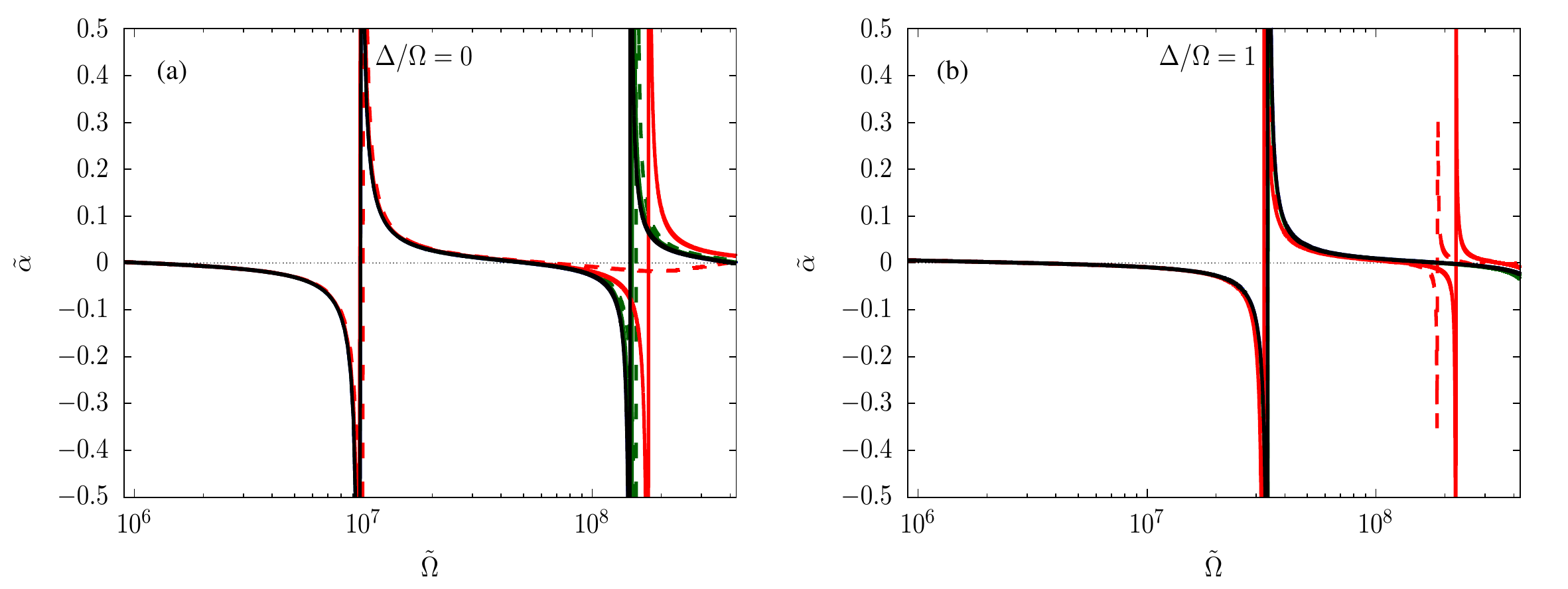}
\caption{Real parts of reduced scattering lengths $\tilde{\alpha}$ at $\tilde{E}_\textrm{coll}=1000$ for NaCs (blue), NaRb (green) and NaK (red), compared to the universal result (black) at (a) $\Delta/\Omega=0$ and (b) $\Delta/\Omega=1$. Solid and dashed lines show results obtained with extended basis sets with $n_\textrm{max}=1$ and $n_\textrm{max}=2$, respectively. In some cases the lines for individual molecules lie underneath the universal result.}
\label{fig:nonuniv-scatlen-nmax2}
\end{figure*}

\subsubsection{Scattering lengths}

Figure \ref{fig:nonuniv-scatlen-nmax2} shows the real part $\tilde{\alpha}$ of the reduced scattering length for the same molecules as Fig.\ \ref{fig:nonuniv-rates} under the same conditions. All the molecules show near-universal behavior for $\tilde\Omega\lesssim 10^8$, which includes both the first pole and the subsequent zero in $\alpha$. NaRb and NaCs remain near-universal until the second pole and beyond. As a consequence, the first bound state will appear at almost the same value of $\tilde{\Omega}$ for all molecules, and its reduced binding energy will be a universal function of $\tilde{\Omega}$ well beyond that.

The scattering length for NaK, which has $\tilde{b}$ only $8.2\times 10^8$, does show strong deviations from universality for $\tilde{\Omega}\gtrsim 10^8$. For NaK, the second pole appears at lower $\tilde{\Omega}$ than universal for $\Delta/\Omega=1$, but in the opposite direction for $\Delta/\Omega=0$; in the latter case, the second pole does not appear at all with $n_\textrm{max}=2$. These deviations for NaK arise because the Rabi frequency in this region is a substantial fraction of its rotational constant; when functions with $n=2$ are included, the wavefunctions for even the isolated molecules include significant contributions from $n=2$ as well as $n=0$ and 1. This significantly alters $d_\textrm{eff}$ and the physics of shielding.

\subsection{Extension to fermionic molecules}
\label{subsec:fermions}

For scattering of identical fermions, the coupled-channel equations include partial-wave functions with $L$ odd instead of even. The rotational pair-state functions are the same, with 10 field-dressed pair states that are near-degenerate at long range and dominate the coupled-channel equations. Different combinations of rotational and partial-wave functions contribute for different values of $M_\textrm{tot}$.

Figure \ref{fig:adiabats-fermions} shows the effective potentials for p-wave scattering ($L=1$) for fermionic Na$^{40}$K and $^{6}$LiCs in the 10-pair basis set; they are different for incoming $M_L = M_\textrm{tot} = -1$, 0 and $+1$, and different from those for bosons, but in reduced units are the same for different fermionic molecules. The scattering properties for fermions are thus universal in the 10-pair basis set, but different from those for bosons.

\begin{figure*}[tb]
\includegraphics[width=2.1\columnwidth]{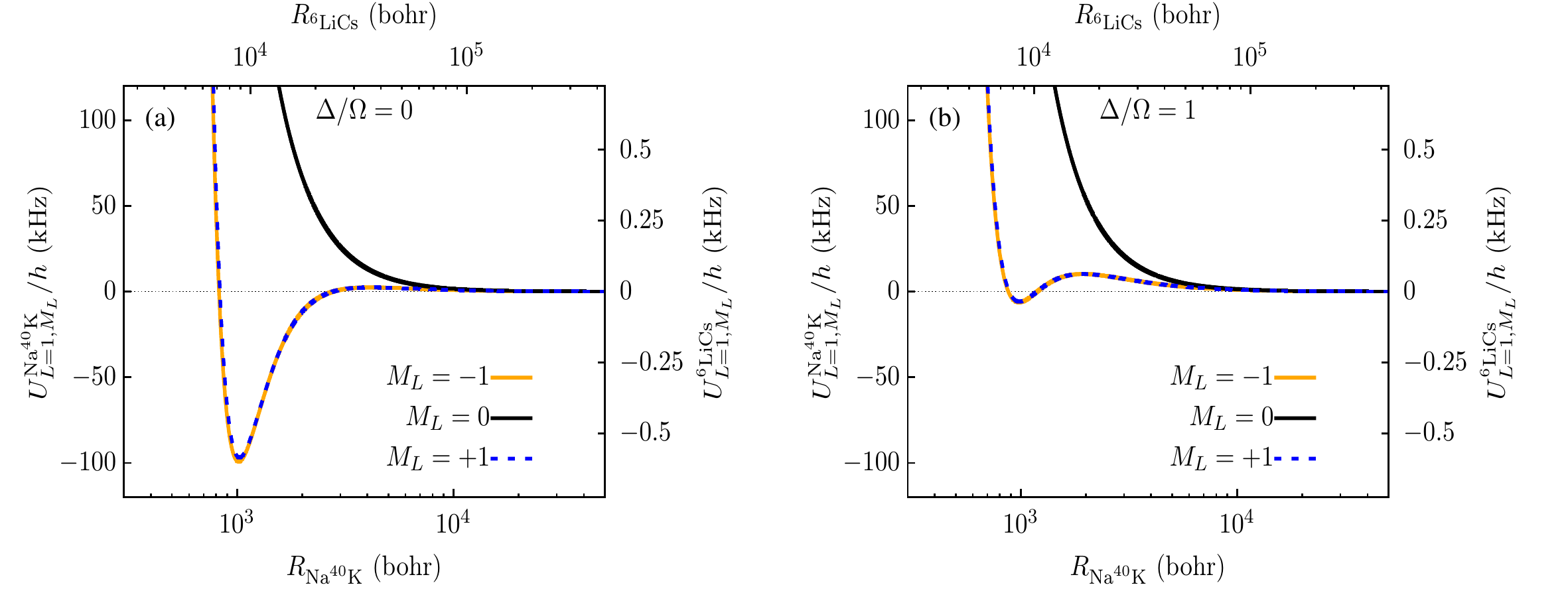}
\caption{Shielding adiabats for p-wave scattering of identical fermions with $\tilde{\Omega} = 1.5\times10^7$, corresponding to $\Omega=50$ MHz for Na$^{40}$K and 0.30 MHz for $^{6}$LiCs, obtained with the 10-pair basis set and demonstrating universal behavior for (a) $\Delta/\Omega=0$; (b) $\Delta/\Omega=1$. There are distinct adiabats for incoming $M_L=-1$, 0 and +1.}
\label{fig:adiabats-fermions}
\end{figure*}

The effective potentials for incoming $L=1$, $M_L=0$ are repulsive at all distances because they are dominated by angles of approach $\theta$ around 0 and $180^\circ$, where the dipole-dipole interaction is repulsive. Those for incoming $M_L=\pm1$ are dominated by $\theta$ around $90^\circ$, where the dipole-dipole interaction is attractive, so they have potential wells outside the shielding repulsion, with centrifugal barriers at long range. The ones for $M_L=-1$ and $+1$ are slightly different because the dark states contribute in different ways. The centrifugal barrier heights are larger for $\Delta/\Omega=1$ than for $\Delta/\Omega=0$ because $d_\textrm{eff}$ is smaller for $\Delta/\Omega=1$.

\subsection{Extension to elliptical polarization}
\label{subsec:ellipt}

For microwaves with elliptical polarization, coupled-channel calculations are more expensive because $M_\textrm{tot}$ is no longer conserved \cite{Karman:shielding-imp:2019}. Nevertheless, microwave shielding is still dominated by basis functions arising from the same 10 pair states as for circular polarization. All other pair states are asymptotically separated from these 10 by energies proportional to $\tilde{b}$ in reduced units. The couplings among the near-resonant set still scale with $E_3$ for fixed ellipticity angle $\xi$. Microwave shielding is thus expected to remain universal, with the same reduced units of length and energy, but with $\xi$ as an additional independent variable.

\subsection{Extension to multiple microwave fields}
\label{subsec:2field}

Experiments on microwave-shielded NaCs \cite{Bigagli:NaCs:2023} and NaRb \cite{Lin:NaRb:2023} have demonstrated the importance of 3-body losses. If uncontrolled, such losses can prevent cooling to quantum degeneracy even when 2-body losses are slow. 3-body losses are dramatically enhanced when the effective potential is deep enough to support 2-molecule bound states, since 3-body recombination then releases kinetic energy and causes trap loss \cite{Stevenson:3body:2025}. As seen above, the first bound state is formed at essentially the same value of reduced Rabi frequency $\tilde{\Omega}$ for all polar molecules.

For NaCs, Bigagli \emph{et al.}\ \cite{Bigagli:BEC:NaCs:2024} suppressed 3-body losses by adding a second microwave field with $\pi$ polarization to supplement the field with $\sigma^+$ polarization. This allowed them to achieve molecular BEC. The $\pi$-polarized field produces a dipole-dipole interaction with the opposite sign to the $\sigma^+$-polarized field, reducing $d_\textrm{eff}$ and the depth of the long-range attractive well. When $d_\textrm{eff}$ is reduced enough that it no longer supports a bound state, 3-body loss is dramatically suppressed.

The $\pi$-polarized field typically has Rabi frequency and detuning that differ from those of the $\sigma^+$-polarized field. Nevertheless, it connects basis functions arising from the same 10 pair states as for circular polarization alone, together with additional near-resonant pair states corresponding to exchange of photons between the two microwave fields \cite{Karman:double:2025}. The couplings among the near-resonant pair states still scale with $E_3$. Microwave shielding will thus again remain universal, with the same reduced units of length and energy, but with additional independent variables to describe the $\pi$-polarized field. This has the important consequence that the methods used to suppress 3-body losses for NaCs \cite{Bigagli:BEC:NaCs:2024} can be transferred directly to other polar molecules, simply by scaling the Rabi frequencies required.

\section{Conclusions}
\label{sec:conc}

We have shown that microwave shielding is near-universal. Different microwave-shielded polar molecules show very similar shielding properties when expressed in suitable reduced units for length and energy. This applies to rate coefficients for elastic scattering and loss, and to scattering lengths. It also applies to the effective potentials (adiabats) that govern shielded collisions and to the properties of the 2-molecule bound states that can exist. It does \emph{not} apply to collisions shielded with static electric fields.

The characteristic length $R_3$ for shielding is proportional to $\mu^2$, where $\mu$ is the molecular dipole moment. The characteristic energy $E_3$ is proportional to $m^{-3}\mu^{-4}$, where $m$ is the molecular mass. The Rabi frequency needed to achieve a particular degree of shielding thus scales with $m^{-3}\mu^{-4}$ and the oscillating electric field $F_\textrm{ac}$ needed to achieve this scales with $m^{-3}\mu^{-5}$. Microwave shielding can be achieved with quite small oscillating fields \cite{Bigagli:NaCs:2023} for molecules such as NaCs, with $\mu=4.75$~D, and is likely to be feasible for most other ultracold polar molecules. However, there are exceptions. For example, RbCs has a dipole moment of only $\mu=1.23$~D \cite{Molony:RbCs:2014}, so would require an oscillating field a factor of 320 larger than for NaCs. This may be too high to be experimentally achievable.

The calculations in the present paper focussed on shielding with a single microwave field of $\sigma^+$ polarization. Nevertheless, universality is expected to extend to elliptical polarization and to situations where two or more microwave fields are used simultaneously, as in recent work to achieve molecular BEC \cite{Bigagli:BEC:NaCs:2024}.

The universality of microwave shielding starts to break down when the Rabi frequency is around 2\% of the molecular rotational constant. For most molecules, this is well beyond the field needed for effective microwave shielding. Among the ultracold molecules of current experimental interest, LiK, LiRb, LiCs, NaK, NaRb, NaCs, KCs, CaF and SrF are all expected to show near-universal behavior at effective shielding fields. However, universality will be less accurate at the Rabi frequencies needed for shielding of LiNa, KRb and RbCs.

An important conclusion of the present work is that the techniques applied to achieve molecular BEC for NaCs \cite{Bigagli:BEC:NaCs:2024}, using two microwave fields of different polarization to suppress 3-body recombination, should be readily transferred to other molecules by scaling the Rabi frequencies appropriately. The transferability is limited only by the ability to achieve the required microwave fields, the different scaled collision energies involved, and the requirement that the Rabi frequencies remain small compared to the molecular rotational constant.

\section*{Rights retention statement}

For the purpose of open access, the authors have applied a Creative Commons Attribution (CC BY) licence to any Author Accepted Manuscript version arising from this submission.

\section*{Data availability statement}

The data that support the findings of this article are openly available from Durham University~\cite{DOI_data-microwave-univ}.

\section*{Acknowledgement}
We are grateful to Ruth Le Sueur, Matthew Frye and James Croft for valuable discussions.
This work was supported by the U.K. Engineering and Physical Sciences Research Council (EPSRC) Grant Nos.\ EP/P01058X/1 and EP/V011677/1.

\bibliographystyle{../long_bib}
\bibliography{../all,MW_shieldingData}
\end{document}